\documentclass[10pt]{article}
\usepackage{amsmath,amssymb,amsfonts}
\usepackage{units}
\usepackage{bbold, lmodern}%opcoes de fonte: mathptmx, helvet, eulervm, avant, fourier, newcent, mathpazo,euler,newcent,sansmath, lmodern
\usepackage{graphicx}% Include figure files
\usepackage{dcolumn}% Align table columns on decimal point
\usepackage{bm,bbm,mathtools,esvect}
\usepackage{slashed}
\usepackage{esvect}
\usepackage{fullpage}
\usepackage[utf8]{inputenc}
\usepackage{mathrsfs}
\title{\vspace{2cm}  \bf CLASSICAL AND QUANTUM \\COMPLEX DYNAMICS \vspace{1cm}}
\author{\bf{SERGIO GIARDINO}\\{\small{\tt sergio.giardino@ufrgs.br}}\\ \vspace{9mm}
\\
\small \sf Departamento de Matem\'atica Pura e Aplicada \\
\small \sf Universidade Federal do Rio Grande do Sul -- UFRGS\\
\small \sf Caixa Postal 15080 -- 91501-970 -- Porto Alegre -- RS \\
\small \sf Brazil}
\begin{document}
\date{} % Remove a data
\maketitle
\newtheorem{theorem}{Theorem}[section] 
\newtheorem{remark}{Remark}[section] 
\newtheorem{lemma}{Lemma}[section] 
\newtheorem{proposition}{Proposition}[section] 
\newtheorem{corollary}{Corollary}[section] 
\newtheorem{definition}{Definition}[section]

\vspace{1cm}
\begin{abstract}
\noindent A generalization of classical mechanics is obtained from a complex parametrization of the phase space. The formalism  supports complex Hamiltonian functions describing non-conservative classical mechanical systems. A quantization scheme that is general enough to incorporate  non-stationary physical processes is also achieved.

\vspace{2mm}
\noindent {\bf keywords:} formalism of classical mechanics; quantum mechanics.

\vspace{1mm}
\noindent {\bf pacs numbers:} 45.20.-d; 03.65.-w.
\end{abstract}

\vspace{5cm}
%\pagebreak

%%%%%%%%%%%%%%%%%%%%%%%%%%%%%%%%%%%%%%%%%%%%%%%%%%%%%%%%%%%%%%%%%%%%%%%%%%
%%%%%%%%%%%%%%%%%%%%%%%%%%%%%%%%%%%%%%%%%%%%%%%%%%%%%%%%%%%%%%%
\section{ Introduction\label{I}}
%%%%%%%%%%%%%%%%%%%%%%%%%%%%%%%%%%%%%%%%%%%%%%%%%%%%%%%%%%%%%%%
%%%%%%%%%%%%%%%%%%%%%%%%%%%%%%%%%%%%%%%%%%%%%%%%%%%%%%%%%%%%%%%%%%%%%%%%%%
The formulation of quantum mechanics in terms of complex numbers represented a rupture with the real formulation of classical mechanics in terms of
real numbers. Even within the symplectic systematization of classical mechanics based on a complex-like structure \cite{Arnold:1989mma,Ratiu:2002isy,Abraham:2008fom}, several differences between the quantum formalism and the classical formalism are understood as analogies whose  meaning remains unexplained.  Well known examples of such kind of parallel involves the classical Poisson bracket and the quantum commutator \cite{Bergmann:1954tc,Franke:1970fs,Millard:1996qj}, as well as the correspondence between the classical canonical transformations and the quantum unitary transformations \cite{Strocchi:1966cca,Kramer:1975gta}.

Nevertheless, puzzling analogies may simply be artificial effects due to the mathematical framework that each theory is based on. Conformal to this hypothesis, a common mathematical language expressing both of the theories may hopefully explain these analogies as particular cases of a certain general theory. The investigation of an unified mathematical  language capable to unify  classical mechanics and quantum mechanics within a single mechanical theory is a matter of research that seems very far from their aim. Among these unifying efforts, \cite{Strocchi:1966cca,Kramer:1975gta} pioneered  a complex statement of classical mechanics that could enable the expression of both of the theories, and a complexification of Hamiltonian dynamics was  examined in \cite{Gerdjikov:2002xd}, as well as \cite{tekkoyun:2006org,cabar:2007clh,Rami:2013iti,castillo:2023app} considered the complexification of the Lagrangian dynamics.  Also partial complex structures, like in the case of the momentum \cite{Albert:2023kah}, were proposed. Likewise, several formulations were tried to unify the whole mechanical theories, such as the usage of differential geometry and Lie structures \cite{Jordan:1961vv}, differential manifolds \cite{Leon:1989nhm,Ashtekar:1997ud,Tekkoyun:2006cch}, hybrid quantum classical dynamics \cite{Prezhdo:1996gs,Braasch:2022hjj}, the quantum Hamilton-Jacobi \cite{Bracken:2003qum,Roncadelli:2007zz,deSouzaDutra:2015mvw,Razavi:2013ito,Carinena:2015drw}, the axiomatization of quantum mechanics as a classical theory  \cite{Heslot:1985qam}, and the quantum-classical dynamics \cite{Jones:1992krw}. In summary, these proposals have advantages and drawbacks, but none of them definitively unified classical and quantum mechanics within a single formalism. Therefore, this issue is still an exciting open question, with many mathematical possibilities, and profound physical implications.

In this article, one presents a complexified classical mechanics, hoping  that it can be useful in a future development of a common language between classical mechanics and quantum mechanics. A differential geometric via treats the curves in the phase space of classical mechanics, relating classical and quantum mechanics by means of a symplectic inner product, relating the curvature of the phase space to the classical energy, the Poisson bracket to the quantum commutator, and the classical complex coordinates to the quantum operators. This apparatus applied to complex Hamiltonian functions describes classical non-linear and non-conservative mechanical systems that can be quantized in a quite natural way. The simple and novel results open several directions of future research, eventually including an unified mechanical theory.

%%%%%%%%%%%%%%%%%%%%%%%%%%%%%%%%%%%%%%%%%%%%%%%%%%%%%
%%%%%%%%%%%%%%%%%%%%%%%%%%%%%%%%%%%%%%%%%%%%%%%%%%%%%
\section{Symplectic classical mechanics\label{MC}}
%%%%%%%%%%%%%%%%%%%%%%%%%%%%%%%%%%%%%%%%%%%%%%%%%%%%%
%%%%%%%%%%%%%%%%%%%%%%%%%%%%%%%%%%%%%%%%%%%%%%%%%%%%%

Before defining the complex phase space, this section resumes the conventional Hamiltonian formalism \cite{Arnold:1989mma,Ratiu:2002isy,Abraham:2008fom} that serves as the prototype to the  formulation  introduced in the sequel. The Hamilton canonical equations satisfied by the dynamical pair composed by the generalized position $q^a$, and the generalized momentum $p_a$ respectively read
\begin{equation}\label{mc01}
 \dot q^a=\frac{\partial H}{\partial p_a},\qquad \mbox{and}\qquad \dot p_a=-\frac{\partial H}{\partial q^a}.
\end{equation}
The dot notation denotes a time derivative, and the dimension index is such that $a=1,\dots d$. The formalism admits a two-dimensional complex formulation. Each dynamical pair associated to a particular  value of the index $a$ admits the definition of the complex variable
\begin{equation}\label{mc002}
 z=\frac{1}{\sqrt{2}}\big( \varkappa_0\, p+iq\big),
\end{equation}
where $\varkappa_0$ is a dimensional parameter to equalize the  dimension of the complex components.  As discussed in the introduction, the idea of writing the Hamilton equations in terms of complex numbers is not new, and a variable like (\ref{mc002}) can be found, for example, in Section 2.1 of \cite{Ratiu:2002isy}. Emphasizing that this idea is motivated as a way to unfold physical relations between classical and quantum mechanics, one also remembers the real character of the classical equations which have to be recovered always when the physical character of the classical motion requires to be examined. Moreover, (\ref{mc002})   enables the equations (\ref{mc01}) to recast as
\begin{equation}\label{mc02}
\dot z=i\varkappa_0 \frac{\partial H}{\partial\bar z},\qquad\mbox{where}\qquad
 \frac{\partial}{\partial \bar z}=\frac{1}{\sqrt{2}}\left(\frac{1}{\varkappa_0}\frac{\partial}{\partial p}+i\frac{\partial}{\partial q}\right),
\end{equation}
where $\bar z$ is the complex conjugate of $z$. 
The complex two-dimensional formalism is thus generalized to $2d$-dimensions using the symplectic matrix protocol \cite{Abraham:2008fom}. Defining the two-dimensional symplectic matrix $J$, 
\begin{equation}\label{mc05}
 J=\left[
\begin{array}{rr}
 0 & 1\\
 -1 & 0
\end{array}
 \right],
\end{equation}
where
\begin{equation}
 J^2=-I
\end{equation}
and  $I$ is the identity matrix, the canonical equations (\ref{mc01}) become
\begin{equation}\label{mc04}
 \bm Z_H=-J \bm dH,
\end{equation}
with
\begin{equation}\label{mc06}
 \bm d H=\left[
 \begin{array}{c}
  H_p \\\varkappa_0 H_q
 \end{array}
 \right],
 \qquad\qquad
 \bm Z_H=
 \left[
 \begin{array}{c}
\varkappa_0  \dot p\\ \dot q
 \end{array}
 \right]
 =\left[
 \begin{array}{c}
  -\varkappa_0 H_q\\ H_p
 \end{array}
 \right],
\end{equation}
where, of course, 
\begin{equation}\label{mc07}
 H_q=\frac{\partial H}{\partial q},\qquad\mbox{and}\qquad H_p=\frac{\partial H}{\partial p}.
\end{equation}
One call $\bm Z_H$ an element of a  real vector space $Z$, and $\bm dH$ an element of the dual space $Z^*$.  This matrix formalism is useful also because it can be easily extended to various dimension through a tensor product. Consequently,
\begin{equation}\label{mc08}
J:Z\to Z^*,\qquad J^{-1}:Z^*\to Z,\qquad\mbox{and}\qquad J^{-1}=J^T=-J.
\end{equation}
By way of clarification, $\bm Z_H$ is slightly different from the notation $\bm X_H$ usually found throughout the literature, and the novel definition (\ref{mc05}) suits to the complex analogy established in the next section.  Finally, the formalism requires the definition of anti-symmetric bilinear forms. Therefore,
\begin{definition}[Symplectic structure]\label{d01} The anti-symmetric bilinear form $\Omega:Z\times Z\to\mathbbm R$
\begin{equation}\nonumber
 \Omega(\bm u,\,\bm v)=\bm u^T J\bm v
\end{equation}
is the symplectic structure within the vector space $Z$.
\end{definition}
And accordingly,
\begin{definition}[Poisson structure]\label{d02} The anti-symmetric bilinear form $\Omega:Z^*\times Z^*\to\mathbbm R$
\begin{equation}\nonumber
 \Omega(\bm\alpha,\,\bm\beta)=\bm\alpha^T J\bm\beta
\end{equation}
is the Poisson structure of the dual space $Z^*$.
\end{definition}
The spaces $Z$ and $Z^*$ are isomorphic, and admit the relation
\begin{equation}\label{mc09}
 \Omega(\bm Z_H,\,\bm u)=-(\bm dH,\,\bm u)
\end{equation}
where $(\bm dH,\,\bm u)$ is the usual inner product. Finally, one defines the Poisson bracket
\begin{definition}[Poisson bracket]\label{d03} Let $F$ and $G$ be real functions. The anti-symmetric bilinear form
 \begin{eqnarray*}
  \{F,\,G\}&\!\!=&\!\!-\Omega(\bm dF,\,\bm dG)\\
  \!\!&=&\!\! \varkappa_0\big(F_q G_p-F_p G_q\big).
 \end{eqnarray*}
 is named the Poisson bracket.
\end{definition}
These definitions subsume the symplectic structure to transpose into a geometric complex structure in the next sections.

%%%%%%%%%%%%%%%%%%%%%%%%%%%%%%%%%%%%%%%%%%%%%%%%%%%%%%%%%%%%%%
\section{Complex phase space \label{CPS}}
%%%%%%%%%%%%%%%%%%%%%%%%%%%%%%%%%%%%%%%%%%%%%%%%%%%%%%%%%%%%%%

In order to describe a phase space of a physical motion parametrized in terms of the complex variable (\ref{mc002}), the complex notation for Hamilton canonical equations (\ref{mc01}) reads
\begin{equation}\label{ce01}
 \dot z=i z_{dH},
\end{equation}
whereby holds the definition
\begin{equation}\label{ce01b}
 z_{dH}=\frac{1}{\sqrt{2}}\Big(H_p+\varkappa_0 H_qi\Big).
\end{equation}
Moreover, from (\ref{mc04}) one obtains 
\begin{equation}\label{ce01a}
\dot z= z_H=iz_{dH},
\end{equation}
what implies $z_H$ and $z_{dH}$ as elements of dual spaces. Considering $\,z(t)\,$ as a plane curve in a complex phase space, the differential geometric formalism \cite{Alencar:2022gpc} suits to describe the system. In order to get it, one first defines the inner product in the complex plane such as
\begin{equation}\label{ce02}
 (z,\,w)=\mathfrak{Re}[z\,\bar w],\qquad\mbox{where}\qquad z,\,w\in\mathbbm C.
\end{equation}
Accordingly, the unitary tangent ($T$) and the normal ($N$) directions are respectively
\begin{equation}\label{ce03}
T=\frac{\dot z}{\left|\dot z\right|},\qquad\mbox{and}\qquad N=i\,T.
\end{equation}
In these circumstances, the complex Frenet equation \cite{Alencar:2022gpc} reads 
\begin{equation}\label{ce04}
\frac{d}{dt} \left(\frac{ \dot z}{\left| \dot z\right|}\right) =i\kappa(t)  \dot z,
\end{equation}
where $\,\kappa(t)\,$ is the curvature real function of the phase space. A little calculation using (\ref{ce02}) gives the curvature function to be
\begin{equation}\label{ce05}
 \kappa=\varkappa_0 \frac{\ddot q\,\dot p-\ddot p\,\dot q}{2\left|\dot z\right|^3}.
\end{equation}
The curve $z$ makes geometrical sense if $q$ and $\varkappa_0 p$ have length dimension ($\lambda$), and thus $\varkappa_0=\tau/\mu$, where one identifies the mass unit ($\mu$) and the time unit ($\tau$). If one imposed a dimensionless $\varkappa_0=1$, a system of units where $\mu=\tau$ holds, and thus time and mass dimensions are equivalent, and also the energy unit ($\epsilon$) is so that 
\begin{equation}\label{ce05a}
\epsilon=\frac{\lambda^2}{\tau}.
\end{equation}
In both of the cases, the dimension unit of the curvature is $1/\lambda$, as expected.  One can identify several correspondences between the symplectic formulation of classical mechanics and complex numbers. In analogy to the canonical symplectic structure of Definition \ref{d01}, the symplectic inner product between $z,\,w\in\mathbbm C$ is such as
\begin{equation}\label{ce13}
 \Omega (z,\,w)=\mathfrak{Re}[z i\bar w]=-\mathfrak{Im}[z \bar w].
\end{equation}
Moreover, remembering the usual inner product (\ref{ce02}), and also in accordance to (\ref{mc09}), it holds that
\begin{equation}\label{ce14}
 \Omega(z_H,\,w)=-(z_{dH},\,w).
\end{equation}
Finally, the complex 
the Poisson bracket
\begin{eqnarray}\label{ce15}
 \nonumber \big\{ F,\,G\big\}&=&-\Omega\big(z_{dF},\,z_{dG}\big)\\
 &=&\varkappa_0\big( F_q G_p-F_p G_q\big)
\end{eqnarray}
keeps perfect accordance to Definition \ref{d03}. 
The inner product likewise satisfies the Jacobi identify
\begin{equation}\label{mc17}
\Omega(u,\,\Omega(z,\,w))+\Omega(z,\,\Omega(w,\,u))+\Omega(w,\,\Omega(u,\,z))=0,
\end{equation}
and therefore the analogy between the complex and the vector formalism is positively established. One can accordingly establish further analogies with quantum mechanics using the complex formalism.

%%%%%%%%%%%%%%%%%%%%%%%%%%%%
\section{Quantization rule}
%%%%%%%%%%%%%%%%%%%%%%%%%%%%%
In an attempt to establish quantum analogies to the complex classical conception of the previous paragraph, one observes that the geometrical structure (\ref{ce03}-\ref{ce05}) alludes to a connection with quantum mechanics to be hypothesized in this paragraph. Taking advantage of a well-known identity easily obtained from the total derivative,  using the classical Hamiltonian one obtains
\begin{equation}\label{ce16}
 \frac{dH}{dt}=\frac{\partial H}{\partial t}+\big\{H,\,H\big\}.
\end{equation}
The conservation of the energy immediately holds for Hamiltonian operators without explicit time dependence, as the Poisson bracket in (\ref{ce16}) is identically zero, and there is no novelty in this. However, a simple quantization scheme can be conjectured taking benefit of the classical harmonic oscillator of mass $m$, elastic constant $k$, and whose Hamiltonian function $H$ meets the usual
\begin{equation}\label{ce07}
H=\frac{p^2}{2 m}+\frac{ kq^2}{2}.
\end{equation}
Therefore,
\begin{equation}\label{ce17}
 z_{dH}=\frac{1}{\sqrt{2}}\left(\frac{p}{m}+\varkappa_0 kqi\right)
\end{equation}
and it holds the identity
\begin{equation}\label{ce18}
\big\{ H,\,H\big\}=-\Omega\big(z_{dH},\,z_{dH}\big)=\varkappa_0\,\omega^2\big(pq-qp\big)
\end{equation}
where
\begin{equation}\label{ce19}
\omega=\sqrt{\frac{k}{ m}}.
\end{equation}
The Poisson bracket (\ref{ce18}) identically zero within a classical regime, confirming the conservative character of the energy of the system already anticipated in (\ref{ce16}). Nevertheless, one can suppose this system to admit a quantum fluctuation of their energy if momentum and position
do not commute anymore. In this situation,  promoting the classical variables to quantum operator, one obtains from (\ref{ce18})
\begin{equation}\label{ce20}
\big\{ H,\,H\big\}\to-\varkappa_0\,\omega^2\big[\hat q,\,\hat p\big]=i\big[\hat z_{dH},\, \hat z^\dagger_{dH}\big],
\end{equation}
and the quantization rule for the harmonic oscillator immediately comes
\begin{equation}\label{ce27}
 \big[\hat z^\dagger_{dH},\, \hat z_{dH}\big]=\varkappa_0\hbar\omega^2.
\end{equation}
One notices $\hat z_{dH}^\dagger$ as the hermitian conjugate of $\hat z_{dH}$, and whose classical analogue is the complex conjugate $\bar z_{dH}$. This is an amusing result, but their real novelty of this quantization scheme seems to be their generality. It is not a particular feature of the harmonic oscillator defined by the Hamiltonian (\ref{ce07}), and can be applied to different physical systems. One also notices that further quantum commutation relation can be obtained. However, it is not so immediate the relation between the classical symplectic product $\Omega$ and the commutation relation. From (\ref{ce13}), one can obtain the classical identity
\begin{equation}\label{ce28}
\frac{i}{2}\Big(\big[z,\,\bar w\big]-\big[\bar z,\, w\big]\Big)=0,
\end{equation}
whose each commutator is of course zero in the case of classical variables, 
but it will be useful when the classical variables where turned into quantum operators. Despite of this, it is possible to obtain quantum commutation relations involving operators obtained from derivatives of $z_{dH}$. For example, using the equations of motion, one obtains the classical result
\begin{equation}
\dot z_{dH}=-\frac{\omega^2}{\sqrt{2}}\Big(q-\varkappa_0 p\,i\Big),
\end{equation}
that, after promoting the position and momentum variables to quantum operators, leads to
\begin{equation}\label{ce24}
	\left[\hat z_{dH},\,\hat{\dot z}^\dagger_{dH}\right]=i\frac{\hbar\,\omega^2}{2}\left(\frac{1}{m}+\varkappa_0^2k\right).
\end{equation}
The interpretation of this result require a geometrical feeling to be developed in the next section.

%%%%%%%%%%%%%%%%%%%%%%%%%%
\section{Geometrical stance}
%%%%%%%%%%%%%%%%%%%%%%%%%%
One can further extend the analogy between quantum and classical mechanics. Using the curvature function of the phase space  graph (\ref{ce05}) and the complex variable (\ref{ce07}), the symplectic inner product (\ref{ce13}) gives
\begin{equation}\label{ce22}
\Omega\big(\dot z_{dH},\ddot z_{dH}\big)=|\dot z_{dH}|^3\omega^2\kappa
\end{equation}
On the other hand the phase space complex function  determined  by the Hamiltonian (\ref{ce01b}) is
\begin{equation}\label{ce07a}
z_{dH}=\frac{\mathcal R\omega}{\sqrt{2}}\Big[ \cos(\omega t)+\varkappa_0 m \omega \sin(\omega t)\,i\Big]
\end{equation}
where $\mathcal R$ is the real amplitude range of the oscillatory motion. One observes that $\varkappa_0=1/m\omega$ would be a suitable choice in this particular case.  However, keeping all the constants, one obtains from the Hamiltonian the classical energy of the system to be
\begin{equation}\label{ce21}
	E=\frac{1}{2}\mathcal R^2 m \omega^2
\end{equation}
Moreover, using (\ref{ce13}) and (\ref{ce21}), one obtains
\begin{equation}\label{ce23}
	\Omega\big(\dot z_{dH},\,\ddot z_{dH}\big)=\frac{\varkappa_0\omega^2}{2}\left(\frac{p^2}{m}+kq^2\right),
\end{equation}
and therefore the classical energy of the system will be the function
\begin{equation}\label{ce29}
E=\frac{\Omega\big(\dot z_{dH},\,\ddot z_{dH}\big)}{\varkappa_0\omega^2}
\end{equation}
thus establishing a relation between the curvature $\kappa$ of the phase space graph (\ref{ce22}) and the energy, namely
\begin{equation}\label{ce25}
	\kappa=\varkappa_0\frac{E}{|\dot z_{dH}|^3}.
\end{equation}
More importantly, the symplectic product (\ref{ce23}) is a way to obtain the classical energy of the system, something that is immediate in the case of elementary Hamiltonian systems, but it is not so obvious in the generalized systems to be developed in the sequel.
Additionally, from (\ref{ce23}) one concludes that 
\begin{equation}
\dot z_{dH} z_{dH}\propto\varkappa_0 E.
\end{equation}
and thus one obtains the quantized energy from (\ref{ce27}) and (\ref{ce24}) to be
\begin{equation}\label{ce26}
\mathcal E=\frac{\hbar\,\omega^2}{2\varkappa_0}\left(\frac{1}{m}+\varkappa_0^2k\right).
\end{equation}
This is a quantum of energy hypothetically related to the curvature $\kappa$ through (\ref{ce25}) and to the amplitude $\mathcal R$ through (\ref{ce21}). Therefore,
if the quantum energy (\ref{ce26}) is a minimum value, this fact can be highlighted using Gromov's non-squeezing theorem \cite{deGosson:2009fop}. This  classical result predicts  a minimum area of the curve determine in the phase space if certain conditions are satisfied. One can thus hypothesize that the quantum energy  (\ref{ce26}) relates to Gromov's minimal classical area, as well as to the curvature of the graph of the motion in the phase space, and one can propose the curvature as a quantization criterion to be further entertained in the next section. To justify initially this appreciation, at least from the classical standpoint, one considers a simple and illuminating example. The dimensional analysis (\ref{ce05a}) in case of $\varkappa_0=1$ enables the identity
\begin{equation}\label{ce08}
E=\frac{\mathcal A}{\mathcal T},\qquad\mbox{where}\qquad \omega \mathcal T=2\pi,
\end{equation}
where the area $\mathcal A$ is generated within the ellipsis (\ref{ce07}) for a particular value $H=E$. Moreover, the  period $\mathcal T$ comprises the motion along the ellipsis with angular velocity $\omega$ determined by (\ref{ce07a}). 

Thus, the symplectic product (\ref{ce22}) relates to the area of the phase space, and may have a minimum value whether a quantization scheme holds. The hypothesis that associates curvature, energy and area will be further emphasized in the next section, and it is also an exciting direction for future research.
The analogy between curvature and area of the phase space is further extended by relation between the complex phase space, and the real phase space verified in Theorem 5.10 of \cite{Alencar:2022gpc}:
\begin{theorem} Let $\mathcal C$ be a regular and convex closed curve $\mathcal C$ of length $\ell$ and area $\mathcal A$. Therefore, it holds that
\[
 \frac{2\pi}{\kappa_m}\leq \ell\leq  \frac{2\pi}{\kappa_M},\qquad\mbox{and}\qquad\frac{\pi}{\kappa_m^2}\leq \mathcal A\leq  \frac{\pi}{\kappa_M^2},
\]
where $\kappa_m$ and $\kappa_M$ are respectively the minimum, and the maximum values of the curvature function $\kappa$ of the curve $\mathcal C$.
\end{theorem}

Remembering the case of a circumference of radius $\mathcal R$, the curvature function is accordingly $\kappa =1/\mathcal R$, and the theorem is immediately verified, as well as in the case of the ellipsis determined by (\ref{ce07}). Moreover, the above theorem confirms that it holds for a circular curve of constant curvature $\kappa$ and length $\ell$ that
\begin{equation}\label{ce09}
2\pi=\omega\mathcal T=\kappa\ell
\end{equation}
indicating a possible quantization rule involving the curvature and a length to be further investigated in future research.
A final and simple application of the geometry of the phase space takes benefit of the derivatives
\begin{equation}\label{ce10}
N=i T, \qquad T_t=|\dot z_{dH}|\kappa N,\qquad N_t=-|\dot z_{dH}|\kappa T,
\end{equation}
as well as 
\begin{equation}\label{ce11}
\dot z_{dH}=|\dot z_{dH}| T,\qquad\mbox{and}\qquad \ddot z_{dH}=\big(|\dot z_{dH}|\big)^{\bm\cdot}\, T+|\dot z_{dH}|^2\kappa N.
\end{equation}
Consequently, the Taylor expansion around $t=t_0$ reads
\begin{equation}\label{ce12}
z_{dH}=(z_{dH})_0+\Delta\big|(\dot z_{dH})_0\big| T_0+\frac{\Delta^2}{2}\Big[\big(\big|(\dot z_{dH})_0\big|\big)^{\bm\cdot} T_0+\big|(\dot z_{dH})_0\big|\kappa_0 N_0\Big]+\mathcal O\left(\Delta^3\right),
\end{equation}
where $\Delta=t-t_0$, and the zero index means the function is evaluated at $t_0$. As expected, a nonzero curvature preserves the normal component at nonlinear terms of the polynomial, and this result will be useful in the next section, in order to relate curvature to quantization.

%%%%%%%%%%%%%%%%%%%%%%%%%%%%%%%%%%%%%%%%%%%%%%%%%%%%%%%%%%
\section{Complex Hamiltonian function \label{CH}}
%%%%%%%%%%%%%%%%%%%%%%%%%%%%%%%%%%%%%%%%%%%%%%%%%%%%%%%%%%
The complex path space formalism introduced in the previous section enables the generalization of the canonical formulation of classical mechanics to include complex Hamiltonian functions. Such a general dynamical theory  is achieved after defining the complex Hamiltonian function
\begin{equation}\label{ch01}
 \mathcal H=H+iK
\end{equation}
where $H$ and $K$ are real functions of $q,\,p$ and $t$. Following (\ref{mc01}-\ref{mc02}), one proposes 
\begin{equation}\label{ch02}
 \dot z=i\varkappa_0\frac{\partial \mathcal H}{\partial\bar z},
\end{equation} 
 and the canonical Hamilton equations accordingly become
\begin{eqnarray}
\nonumber &&\dot q=H_p-\varkappa_0 K_q\\
\label{ch03}  &&\dot p=-H_q-\frac{1}{\varkappa_0}K_p.
\end{eqnarray}
In analogy to (\ref{ce01a}), it holds that
\begin{equation}\label{ch04}
 z_{\mathcal H}=i z_{d\mathcal H}
\end{equation}
and that
\begin{equation}\label{ch04a}
  z_{d\mathcal H}=\frac{1}{\sqrt{2}}\Big[H_p-\varkappa_0 K_q+\big(\varkappa_0 H_q+K_p\big)i\Big]
\end{equation}
Therefore, every formal result of the previous section also hold in the case for complex Hamiltonians, the only differences to be observed are the
components of $z_{d\mathcal H}$, which recover $z_{dH}$ for constant $K$, as expected. Two simple examples will shed some light on the physical character of the imaginary component of the complex Hamiltonian (\ref{ch01}). 
%%%%%%%%%%%%%%%%%%%%%%%%%%%%%%%%%%%%%%%%%%%%%%%%%%%%%%
\paragraph{\underline{Imaginary harmonic oscillator}}
%%%%%%%%%%%%%%%%%%%%%%%%%%%%%%%%%%%%%%%%%%%%%%%%%%%%%%
The pure imaginary Hamiltonian accordingly reads
\begin{equation}\label{ch05}
 \mathcal H=\alpha_0\left(\frac{p^2}{2 m}+\frac{ kq^2}{2}\right)i
\end{equation}
where $\alpha_0$ is a dimensionless real  constant. The equations of motion obtained from (\ref{ch03}) read
\begin{equation}\label{ch05a}
\dot q=-\alpha_0\varkappa_0 kq,\qquad\mbox{and}\qquad\dot p=-\frac{\alpha_0}{m\varkappa_0}p,
\end{equation}
and their solutions immediately obtained encompass
\begin{equation}\label{ch06}
 q(t)=\mathcal R_1 \exp\Big[-\alpha_0\varkappa_0 k t\Big]\qquad p(t)=\mathcal R_1\exp\Big[-\frac{\alpha_0}{m\varkappa_0}t\Big].
\end{equation}
where $\mathcal R_0$ and $\mathcal R_1$ are real amplitudes. One observes decaying or forcing processes from the imaginary dynamics, depending on the signal of the $\alpha_0$ constant.  Therefore,
\begin{equation}
 z_{d\mathcal H}=\frac{\alpha_0}{\sqrt{2}}\left(-\varkappa_0 kq+\frac{p}{m}i\right).
\end{equation}
The classical energy obtained from (\ref{ce29}) to be
\begin{equation}
E=\frac{\alpha_0^5\omega^2}{4\varkappa_0}\left(\frac{1}{m}-\varkappa_0^2k\right)pq,
\end{equation}
demonstrating that the classical energy decays, because the process a dissipative carries out.
The quantization technique developed in the previous sections enables to obtain
\begin{equation}\label{ch23}
	\big[\hat z^\dagger_{dH},\, \hat z_{dH}\big]=\alpha_0^2\varkappa_0\hbar\omega^2.
\end{equation}
And the correspondence to the usual harmonic oscillator (\ref{ce27}), even though the physical character is different. In order to obtain the quantum energy, one considers
\begin{equation}\label{ce30}
	\big[\hat{\dot z}^\dagger_{dH},\, \hat{\ddot z}_{dH}\big]=\frac{\alpha_0^5\hbar\omega^4}{2}\left(\frac{1}{m}+\varkappa_0^2k\right)
\end{equation}
and thus
\begin{equation}
	\mathcal E=\alpha_0^5\frac{\hbar\,\omega^2}{2\varkappa_0}\left(\frac{1}{m}+\varkappa_0^2k\right),
\end{equation}
and another perfect agreement to the previous case (\ref{ce26}).

Finally, a comment on the condition
\begin{equation}\label{ch08}
km=\frac{1}{\varkappa_0^2}
\end{equation}
implying in (\ref{ch05a}-\ref{ch06}) that
\begin{equation}\label{ch09}
 p\propto q.
\end{equation}
There are several consequences of this condition. First of all (\ref{ce05}) is identically zero, meaning that the physical motion (\ref{ch05a}) represents  a straight line in the phase space, whose curvature is of course zero. Additionally, (\ref{ch09}) would imply that
\begin{equation}
 [\hat q,\,\hat p]=0.
\end{equation}
Hence one has another evidence for an association between the curvature of the classical phase space and the quantization of the corresponding physical motion, where the points of zero curvature of the phase space corresponds to non-quantizable physical situations. This is an exciting direction for future investigation.
%%%%%%%%%%%%%%%%%%%%%%%%%%%%%%%%%%%%%%%%%%%%%%%
\paragraph{\underline{Attenuated harmonic oscillator}}
%%%%%%%%%%%%%%%%%%%%%%%%%%%%%%%%%%%%%%%%%%%%%%%
By way of a second example, one  considers the complex Hamiltonian function
\begin{equation}\label{ch10}
\mathcal H=\frac{p^2}{2 m}+\frac{ kq^2}{2}+i\frac{\beta_0}{n+1}p^{n+1}
\end{equation}
where $\beta_0$ is a real constant. Therefore, (\ref{ch04a}) renders
\begin{equation}\label{ch11}
 z_{d\mathcal H}=\frac{1}{\sqrt{2}}\left[\frac{p}{m}+\big(\varkappa_0 kq+\beta_0p^n\big)i\right],
\end{equation}
and the equations of motion accordingly read
\begin{eqnarray}
\nonumber && p=m\dot q\\
\label{ch12} && \dot p=-kq-\frac{\beta_0}{\varkappa_0} p^n.
\end{eqnarray}
The physical system admits a self-interaction of the system determined by the $p^n$ term. Therefore,
\begin{equation}\label{ch13}
 \ddot q+m^{n-1}\frac{\beta_0}{\varkappa_0}\dot q^n+\omega^2q=0.
\end{equation}\label{ch14}
The particular case of $n=1$ provides the simplest solutions. Defining
\begin{equation}\label{ch15}
\Lambda^2=\left(\frac{\beta_0}{2\varkappa_0}\right)^2-\omega^2,
\end{equation}
one obtains
\begin{equation}\label{ch16}
\left\{
\begin{array}{ll}
 q(t)=\mathcal R_0 \exp\big[k_+t\big]+\mathcal S_0 \exp\big[k_-t\big] & \Lambda^2>0\\ \\
 q(t)=\Big(\mathcal R_0+\mathcal S_0 t\Big)\exp\left[\frac{\beta_0}{2\varkappa_0}t\right]&\Lambda^2=0\\ \\
 q(t)=\Big(\mathcal R_0\sin\Lambda t+\mathcal S_0\cos\Lambda t\Big)\exp\left[\frac{\beta_0}{2\varkappa_0}t\right]&\Lambda^2<0,
\end{array}
\right.
\end{equation}
where
\begin{equation}\label{ch17}
 k_a=\frac{\beta_0}{2\varkappa_0}\pm\Lambda,\qquad \mbox{with}\qquad a=\big\{+,\,-\big\}.
\end{equation}
Of course, $\lambda^2\geq0$ encompasses non-oscillatory processes, while $\Lambda<0$ is related to oscillatory processes. Using the equations of motion to eliminate the time derivatives, from (\ref{ch11}) one obtains
\begin{eqnarray}
\label{ch18} &&\dot z_{d\mathcal H}=\frac{1}{\sqrt{2}}\left[\,-\omega^2q-\frac{\beta_0}{m\varkappa_0}p+\varkappa_0\left[\left(\omega^2-\frac{\beta_0^2}{\varkappa_0^2}\right)p-\frac{\beta_0k}{\varkappa_0}q\right]i\,\right],\\
\label{ch19} &&\ddot z_{d\mathcal H}=\frac{1}{\sqrt{2}}\left[\,\frac{1}{m}\left(\frac{\beta_0^2}{\varkappa_0^2}-\omega^2\right)p+\frac{\beta_0\omega^2}{\varkappa_0}q+\varkappa_0\Big[\frac{\beta_0}{\varkappa_0}\left(\frac{\beta_0^2}{\varkappa_0^2}-2\omega^2\right)p
 +\left(\frac{\beta_0^2}{\varkappa_0^2}-\omega^2\right)kq\Big]i\,\right].
\end{eqnarray}
In this case, the classical energy obtained following the same procedure used in (\ref{ce29}) reads,
\begin{equation} \label{ch20}
 \Omega\big(\ddot z_{d\mathcal H},\,\dot z_{d\mathcal H}\big)=\frac{1}{2}\left(\frac{p^2}{m}+kq^2+\frac{\beta_0}{\varkappa_0}	 pq\right),
\end{equation}
The quantum commutation (\ref{ce27}) is recovered, and 
\begin{equation}\label{ch21}
 \Big[\hat{\dot z}_{d\mathcal H},\,\hat{\ddot z}_{d\mathcal H}^\dagger\Big]=i\hbar\left[\frac{\omega^4}{2}\left(\frac{1}{m}+\varkappa_0^2k\right)+\frac{k\beta_0^4}{\varkappa_0^2}\right] -\beta_0\hbar\omega^2\left(\frac{\omega^2}{2}+\frac{\beta_0^2}{\varkappa_0^2}\right),
\end{equation}
and the quantum energy will be
\begin{equation}
\mathcal E=\hbar\left[\frac{\omega^2}{2\varkappa_0}\left(\frac{1}{m}+\varkappa_0^2k\right)+\frac{k\beta_0^4}{\omega^2\varkappa_0^3}\right]
\end{equation}
what establishes the quantization of a non-stationary process, a remarkable result, because (\ref{ch21}) generalizes  (\ref{ce24}) and (\ref{ce26}), recovering the previous result in the particular case for $\beta_0=0$, as expected.  It also demonstrates that the real component of the Hamiltonian (\ref{ch10}) describes the oscillatory motion, and is responsible for the imaginary term of (\ref{ch21}), whereas the imaginary component of (\ref{ch10}) classically generates a non-oscillatory motion that engenders the real contribution to (\ref{ch21}). Finally, it can be observed that the choice
\begin{equation}\label{ch22}
 p=\pm\sqrt{km}q,
\end{equation}
 as well as a certain choice of $\beta_0$ and $\varkappa_0$, sets to zero the right hand side of (\ref{ch21}) in accordance to (\ref{ch09}), although it does not set to zero the right hand side of (\ref{ch20}). This is another consistency test of the theory, indicating (\ref{ch09}) as a characteristic condition of non-quantizable dynamics.

 One immediately observes the right hand  side of (\ref{ch21})to be a complex number, a clear difference from an imaginary number appearing in (\ref{ce24}). This fact can be understood as a sign of the non-oscillatory character of the classical solution. Differences involving imaginary and real results of quantum commutators have already been observed in the context of general imaginary units \cite{Giardino:2023uzp}, but such a complex Hamiltonian represents a more general result, that can encompass the quantization of either forced or attenuated processes, confirming previous quantum studies on dissipative systems \cite{Kanai:1948dfb}. Finally, these examples show the complex formalism as a possible way to represent dissipative physical processes, and this is an exciting novel direction for research.

%%%%%%%%%%%%%%%%%%%%%%%%%%%%%%%%%%%%%%%%%%%%%%%%%%%%%%%%%%%%%%%%%%%%%%%%%%
%%%%%%%%%%%%%%%%%%%%%%%%%%%%%%%%%%%%%%%%%%%%%%%%%%%%%%%%%%%%%%%
\section{ Conclusion \label{C}}
%%%%%%%%%%%%%%%%%%%%%%%%%%%%%%%%%%%%%%%%%%%%%%%%%%%%%%%%%%%%%%%
%%%%%%%%%%%%%%%%%%%%%%%%%%%%%%%%%%%%%%%%%%%%%%%%%%%%%%%%%%%%%%%%%%%%%%%%%%
In this article, one examined several original aspects of classical mechanics, as well as their relation to quantum mechanics. From the classical point of view,  the complex parametrization for the phase space  described in Section \ref{CPS} enabled the consideration of complex Hamiltonian functions. From the quantum standpoint, the ordinary relation between the Poisson bracket and the quantum commutator  (\ref{ce20}) has been established in a untried way throughout the symplectic inner product of the complex plane  (\ref{ce02}). This quantization rule supports non-stationary physical processes,  such that described in terms of complex Hamiltonian functions in Section \ref{CH}, what is an interesting find. Another aspect of the  rules shows the real components of the Hamiltonian functions  as responsible for imaginary contributions to the quantum commutator  (\ref{ch21}), whereas the imaginary components contribute to the real elements of the quantum commutator. The results are also important because they provide a way to study dissipative physical processes, whose Hamiltonian and Lagrangian formalism are still being developed.

The simpleness of these results asserts their  possible importance at a first sight, what opens numerous  alternatives for future investigation. These include the relation between the curvature of the path within the complex phase space and the quantization rule, generalization to higher dimensional phase space \cite{encheva:2009rpe}, the geometry of dissipative systems \cite{CIAGLIA2018159} the research of classical solutions for complex Hamiltonian functions, complex particles \cite{Hori:2024vly}, quantum computation \cite{Volovich:2024nzw}, creation-annihilation processes \cite{Pogrebkov:2024kbg}, the investigation of dissipative as well as non-linear classical and quantum systems, and the formulation of classical and quantum field theories, to cite only a few.

\paragraph{Funding} This work is supported by the Funda\c c\~ao de Amparo \`a Pesquisa do Rio Grande do Sul, FAPERGS, grant number 23/2551-0000935-8 within the Edital 14/2022.

\paragraph{Data availability statement} The author declares that data sharing is not applicable to this article as no data sets were generated or analyzed during the current study.

\paragraph{Declaration of interest statement} The author declares that he has no known competing financial interests or personal relationships that
could have appeared to influence the work reported in this paper.
    
%%%%%%%%%%%%%%%%%%%%%%
%
%
%  BIBLIOGRAPHY
%
%
\begin{footnotesize}
\bibliographystyle{unsrt} 
\bibliography{bib_MCH_1}
\end{footnotesize}
\end{document}